\documentclass[twocolumn,showpacs,preprintnumbers,prb]{revtex4-1}
\usepackage{eurosym}
\usepackage{graphicx,bm,amsmath,amssymb}

\def\gz{\ifmmode{Z\hskip -4.8pt Z}
    \else{\hbox{$Z\hskip -4.8pt Z$}}\fi}

\newcommand{\be}{\begin{equation}}
\newcommand{\ee}{\end{equation}}
\newcommand{\bea}{\begin{eqnarray}}
\newcommand{\eea}{\end{eqnarray}}

\begin{document}

\title{Phase diagram of the ionic Hubbard model with 
density-dependent hopping}

\author{P. Roura-Bas}
\affiliation{Centro At\'{o}mico Bariloche, GAIDI, CONICET,  8400 Bariloche, Argentina}

\author{A. A. Aligia}
\affiliation{Instituto de Nanociencia y Nanotecnolog\'{\i}a CNEA-CONICET, GAIDI,
Centro At\'{o}mico Bariloche and Instituto Balseiro, 8400 Bariloche, Argentina}

\begin{abstract}
We obtain the quantum phase diagram of the ionic Hubbard 
model including electron-hole symmetric density-dependent hopping.
The boundaries of the phases are determined by crossing of excited levels with particular discrete symmetries, which coincide with jumps of charge and spin Berry phases with a topological 
meaning. Reducing the magnitude of the hopping terms that do not
change the total number of singly occupied sites with respect to the other one, the region of the phase diagram
occupied by the fully gapped spontaneously dimerized insulator (which separates
the band insulating and Mott insulating phases) is enlarged,
particularly for small values of the alternating on-site energy.
This result might be relevant for experiments in cold atoms in which topological charge pumping is observed when alternation 
in the hopping is included.

\end{abstract}



\maketitle

\section{Introduction}

\label{intro}

Ultracold quantum gases provide a versatile platform as universal
quantum simulators of many-body problems \cite{gross}. Cold atoms
as well as other platforms have been used to study quantized
topological charge pumping in driven systems \cite{citro}.
A time dependent adiabatic evolution in a closed cycle in a certain space of parameters constitute a Thouless pump, in which a quantized amount of charge or spin is transported, which is topologically protected \cite{thou,niu}.  
Simulating the non-interacting
Rice-Mele model (RMM) \cite{rice} with ultracold atoms, quantized
charge pumping has been achieved for bosons \cite{nakaji} 
and fermions \cite{loh}. More recently, charge pumping in the 
fermionic interacting RMM (IRMM) has been studied experimentally \cite{konrad} and theoretically \cite{nakag,eric},
including spin pumping \cite{eric}.

The IRMM is a one-dimensional Hubbard model which includes alternating on-site energies $\pm \Delta$ and hopping $t \pm \delta$ 
[Eq. (\ref{hirm}) with $t_{\alpha \beta}=t$].
Ideally, in a Thouless pump, a critical point of degeneracy is 
surrounded in the adiabatic time cycle without closing a gap.
For the IRMM, and fixed on-site interaction $U$, the 
cycles which lead to non-trivial charge pumping enclose 
critical points lying at $\delta=0$ and $\Delta=\pm \Delta_c$.
For $\delta=0$, the IRMM is equivalent to the ionic Hubbard 
model (IHM) \cite{sdi,phihm,manih,rihm,abn,stenzel} and 
it is known that at $\Delta=\pm \Delta_c$ there is a charge
transition in which the topologically protected charge 
Berry phase jumps between the values $0$ and $\pi$ \cite{phihm},
implying a transport of one charge when a time cycle is performed
in the plane ($\delta, \Delta$) enclosing the point 
($0, \pm \Delta_c$) \cite{eric,zup}. 

Similarly at $\Delta=\pm \Delta_s$ there is a spin transition in 
the IHM ($\delta=0$)
with a jump in the spin Berry phase and a closing of the spin gap 
for $|\Delta| \leq \Delta_s < \Delta_c$. The IHM has three phases.
The system is a Mott insulator (MI) for $0 \leq \Delta < \Delta_s$, a band insulator (BI) for $\Delta > \Delta_c$ and a spontaneously
dimerized insulator (SDI) in a narrow region between the other two.
The phase diagram (which is symmetric by a change of sign in
$\Delta$) has been constructed in Ref. \onlinecite{phihm} using 
the method of crossing of excited energy levels (MCEL) based on conformal field theory \cite{naka2,nomu,naka,naka1,som}.
The spin gap opens as $|\delta|^{2/3}$ leading to a dimerized 
phase for finite $\delta$ \cite{eric}.

The fact that the spin gap vanishes in the MI phase 
(that corresponds to $\delta=0$, $-\Delta_s < \Delta < \Delta_s$)
brings problems for the charge pumping. A cycle in the plane 
$\delta, \Delta$ that encloses a critical point $\Delta=\pm \Delta_c$ and passes far from it, necessarily traverses the MI 
phase, because $\Delta_c$ and $\Delta_s$ are very near each other.
Traversing with a finite velocity a gapless point produces 
spin excitation at finite energy, which in turn lead to charge excitations because of the mixing of both sectors at finite energy 
\cite{sdi,rihm}. This leads to oscillations in the charge pumping
and loss of quantization with the number of cycles as determined theoretically \cite{eric} and experimentally \cite{expe}. 
While addition of a staggered magnetic field or Ising spin
interactions lead to opening of the spin gap and robust charge pumping \cite{eric}, these terms are not experimentally
feasible at present.

Another possibility to enlarge the spin gap and the region of
stability of the SDI phase is to add a 
density-dependent hopping (DDH).
Such a term in an electron-hole symmetric form has been realized
in cold atoms using Floquet engineering \cite{ma,meine,desbu,go1,messer,goerg}. The Hubbard model with 
nearest-neighbor hopping dependent on the occupancy of the
sites involved (also called correlated hopping) 
has been derived and studied as an effective model for the superconducting cuprates \cite{fedro,sim,brt}, 
which leads to enhancement of superconductivity for certain 
parameters \cite{hir,hir2,lili,jiang}. In one dimension, it 
has been found that when the hopping term that changes the 
number of singly occupied sites [$t_{AB}$ in Eq. (\ref{hirm})] is larger
than the other two, a dimerized phase with a spin gap is favored \cite{jaka,bosolili,phtopo,naka2}, which is the desired effect. 

The goal of this work is to study to what extent the region of 
the phase diagram of the IHM occupied by the fully gapped 
SDI phase can be enlarged including DDH.
We use the method of crossing of excited energy 
levels (MCEL) based on conformal field theory 
\cite{naka2,nomu,naka,naka1,som}, already used in 
Ref. \onlinecite{phihm} for the standard IHM. 
For this model including DDH, the method also coincides
with that of jumps of charge and spin Berry phases used in 
Ref. \onlinecite{phtopo}.

The paper is organized as follows. In Section \ref{model} we explain the model and methods. The resulting phase diagram is contained in Section \ref{res}. Section \ref{sum} contains a summary and discussion.

\section{Model and methods}

\label{model}

The IRMM including DDH has the form

\begin{eqnarray}
H &=&\sum_{j \sigma} \left[ -1+\delta \;(-1)^{j}\right]
 \left( c_{j\sigma }^{\dagger }c_{j+1\sigma
}+\text{H.c.}\right)   \notag \\
&&\times [t_{AA}(1-n_{j\bar{\sigma}})(1-n_{j+1\bar{\sigma}}) 
+t_{BB}n_{j\bar{\sigma}}n_{j+1\bar{\sigma}}
\notag \\
&&+t_{AB}(n_{j\bar{\sigma}}
+n_{j+1\bar{\sigma}}-2n_{j\bar{\sigma}}n_{j+1\bar{\sigma}})]  \notag \\
&&+\Delta \sum_{j \sigma}(-1)^{j}n_{j\sigma }
+U\sum_j n_{j\uparrow }n_{j\downarrow }.  \label{hirm}
\end{eqnarray}

The first term is the DDH, which is alternating 
for $\delta \neq 0$. The amplitude $t_{AA}$ corresponds
to the situation in which only the particle that hops
occupies the two nearest-neighbor sites involved in
the hopping. For $t_{AB}$ and $t_{BB}$ the total occupancy
is 2 and 3 respectively. In the following we assume the 
electron-hole symmetric case $t_{BB}=t_{AA}$, which is the 
one implemented experimentally with cold atoms \cite{ma,meine,desbu,go1,messer,goerg}.
$\Delta$ is the alternating on-site energy and $U$ is the 
on-site Coulomb repulsion, both characteristic of the 
IHM \cite{sdi,phihm,manih,rihm,abn,stenzel}.

Our conclusions, and our discussions below on the effect 
of the alternation of the hopping $\delta$ are the same 
if $\delta$ affects only the hopping part proportional
to $t_{AB}$, and not the other two.

In experiment usually the pump cycles are done in a two-dimensional
space $(\delta,p)$ in which both $\delta$ and another parameter 
$p$ (like $\Delta$ or $U$) depend on time and return to the original value after the cycle. In the adiabatic limit, 
the charge (spin) pumped in the cycle is determined by the evolution of the charge (spin) Berry phase $\gamma_c$ ($\gamma_s$) in the cycle (see for example Ref. \onlinecite{eric}). 
Non trivial quantized charge (spin) pumping takes place when
a critical point at which $\gamma_c$ ($\gamma_s$) jumps,
is surrounded in the cycle.
The critical points lie on the line $\delta=0$, because 
for $\delta=0$, the system has inversion symmetry at each site 
and as a consequence, the Berry phases can only be either
0 or $\pi$ (mod $2\pi$). In other words, $\gamma_c/\pi$ and $\gamma_s/\pi$
become topological numbers protected by inversion symmetry \cite{zup}. In addition, the MI phase in which the spin gap vanishes is also restricted to $\delta=0$. Then to identify 
a possible cycle that encloses the charge critical point
with a ground state separated from the rest of the spectrum
in the whole cycle, one can keep $\delta=0$, where all ground-state 
degeneracies lie. This is what we 
do in the rest of the work. The model becomes the 
IHM with electron-hole symmetric DDH.

To calculate the phase diagram of the model we use the 
MCEL \cite{nomu,naka,naka1,naka2,som}. The idea of the method 
is that the dominant correlations at large distances correspond to the smallest excitation energies. The crossing of excited levels 
in different symmetry sectors therefore correspond to phase 
transitions. The method has been used before for similar 
models \cite{phihm,naka2,phtopo}. For our model, this method and the jumps in the values of the Berry phases give the same information \cite{phihm}, but the MCEL is easier to implement.

The crossing for both charge and spin transitions are determined 
using open-shell boundary conditions (periodic if the 
number of sites $L$ is multiple of 4, antiperiodic for
$L$ even not multiple of 4). The charge transition is determined
by a crossing in the ground state of the two singlets of 
lowest energy with opposite
parity under inversion. In the BI phase the ground state is even under inversion, while it is odd in the other two phases. 
The spin transition between SDI and MI phases, is determined by 
the crossing of the excited even singlet with lowest energy 
and the lowest excited odd triplet, which has less energy in the MI phase. In the actual calculation we have not evaluated the total spin $S$ of the states, but used the parity under time reversal (the singlet is even and the triplet with total spin projection 
$S_z=0$ is odd). All  these states have wave vector 0 for $\Delta \neq 0$.

\begin{figure}[h]
\begin{center}
\includegraphics[width=7cm]{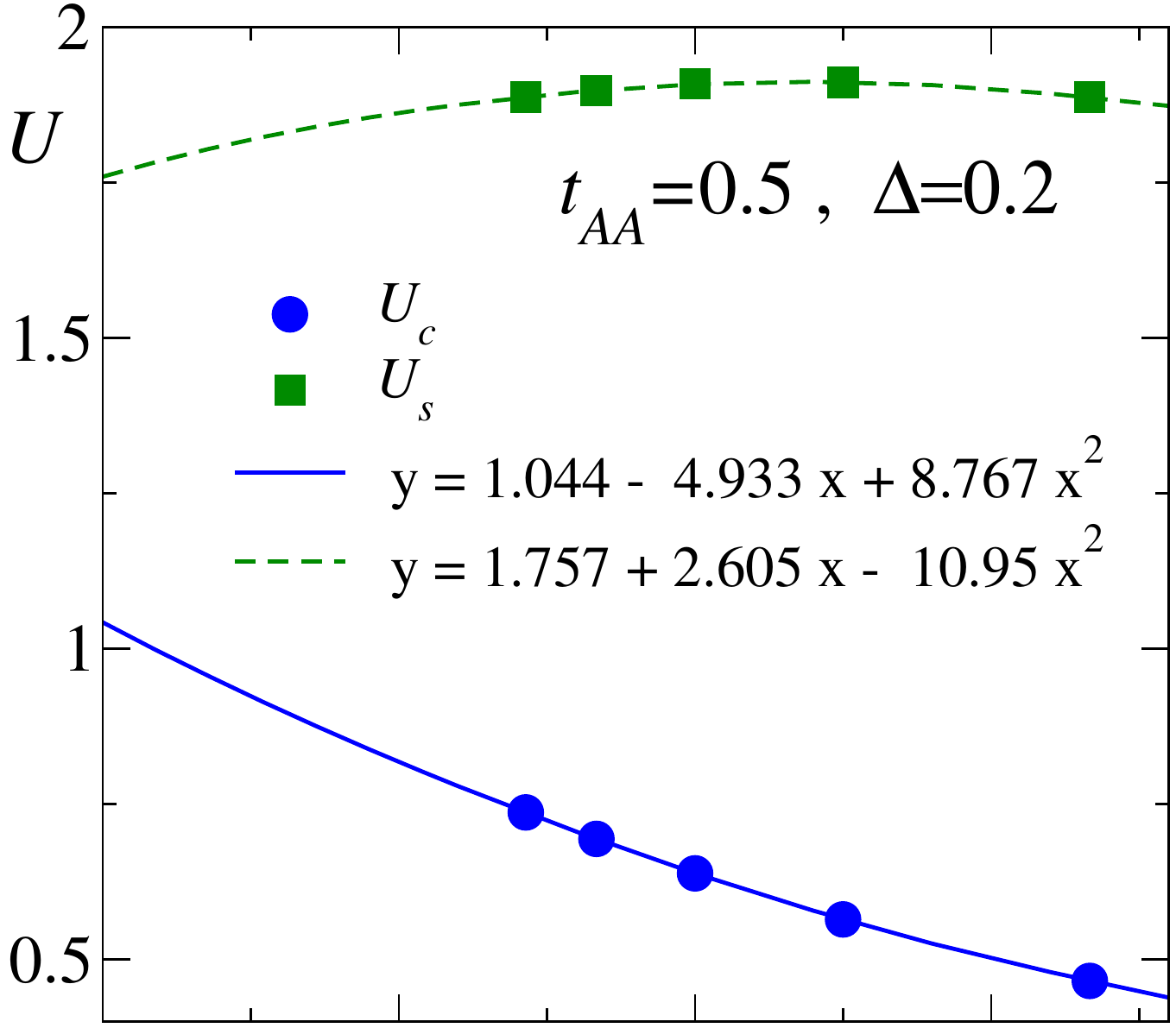}
\includegraphics[width=7cm]{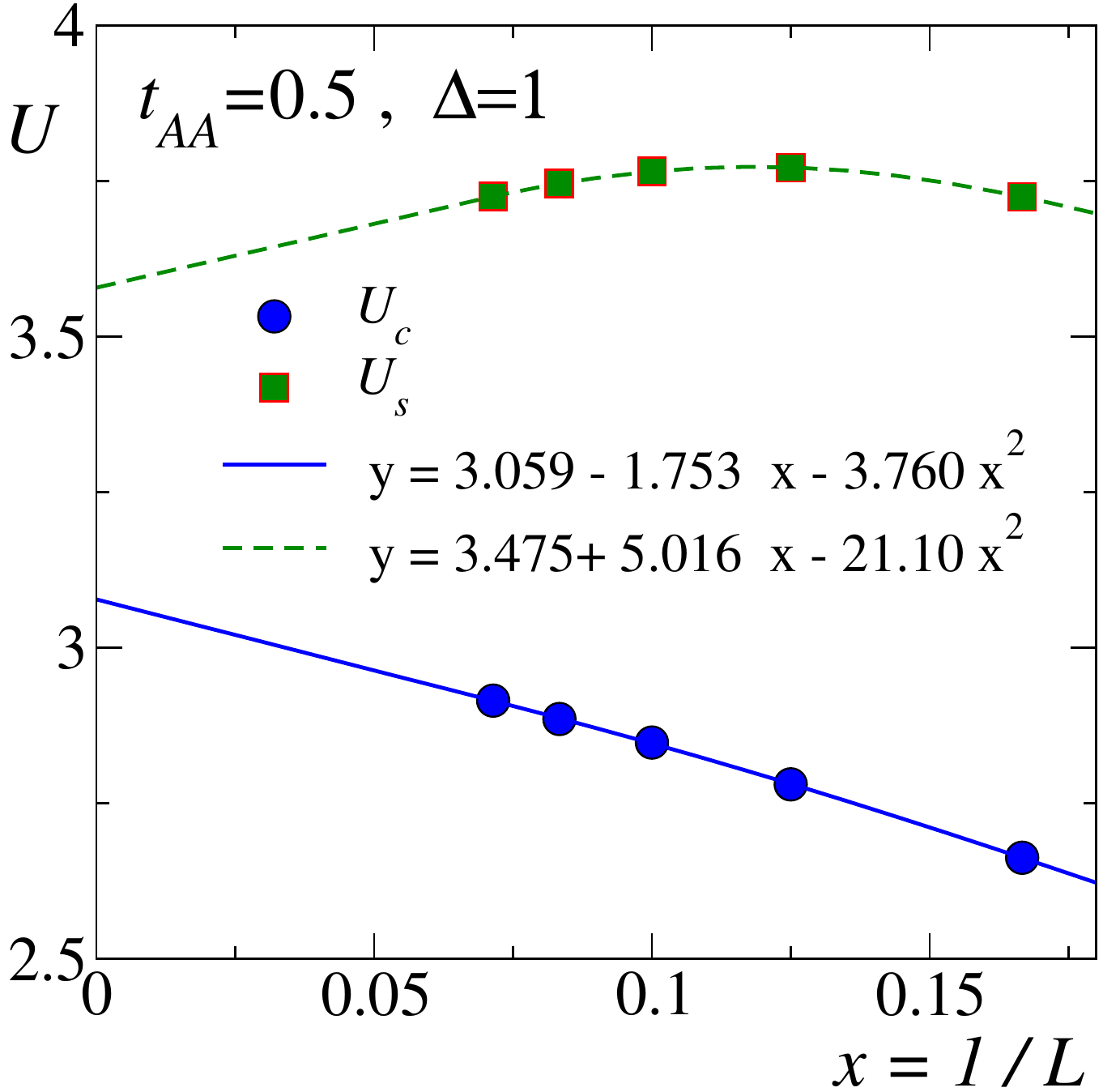}
\caption{(Color online) Critical values of $U$ 
for the charge and spin 
transitions for $t_{AB}=1$, and other parameters indicated 
inside each figure.}
\label{extrap}
\end{center}
\end{figure}

To determine the phase diagram we have set $t_{AB}=1$ as the unit of energy. Then for a given value of $t_{AA}$ and $\Delta$ we have calculated the values of $U$ that correspond to the charge
($U_c$) and spin ($U_s$) transitions using the MCEL for all 
even number of sites $L$ in the range $6 \leq L \leq 14$. The results were extrapolated to $L \rightarrow \infty$ using a 
quadratic polynomial in $1/L$. Examples of the extrapolation 
are shown in Fig. \ref{extrap}. The curves fits well the data 
and the finite-size effects are in general small, except for the charge transition for small values of $\Delta$. In any case, a deviation of the value of $U_c$ for $\Delta=0.2$ for up to
20\%  is very unlikely from the trend of the curve and does not modify our conclusions.

\section{Results}

\label{res} 

\begin{figure}[h]
\begin{center}
\includegraphics[width=\columnwidth]{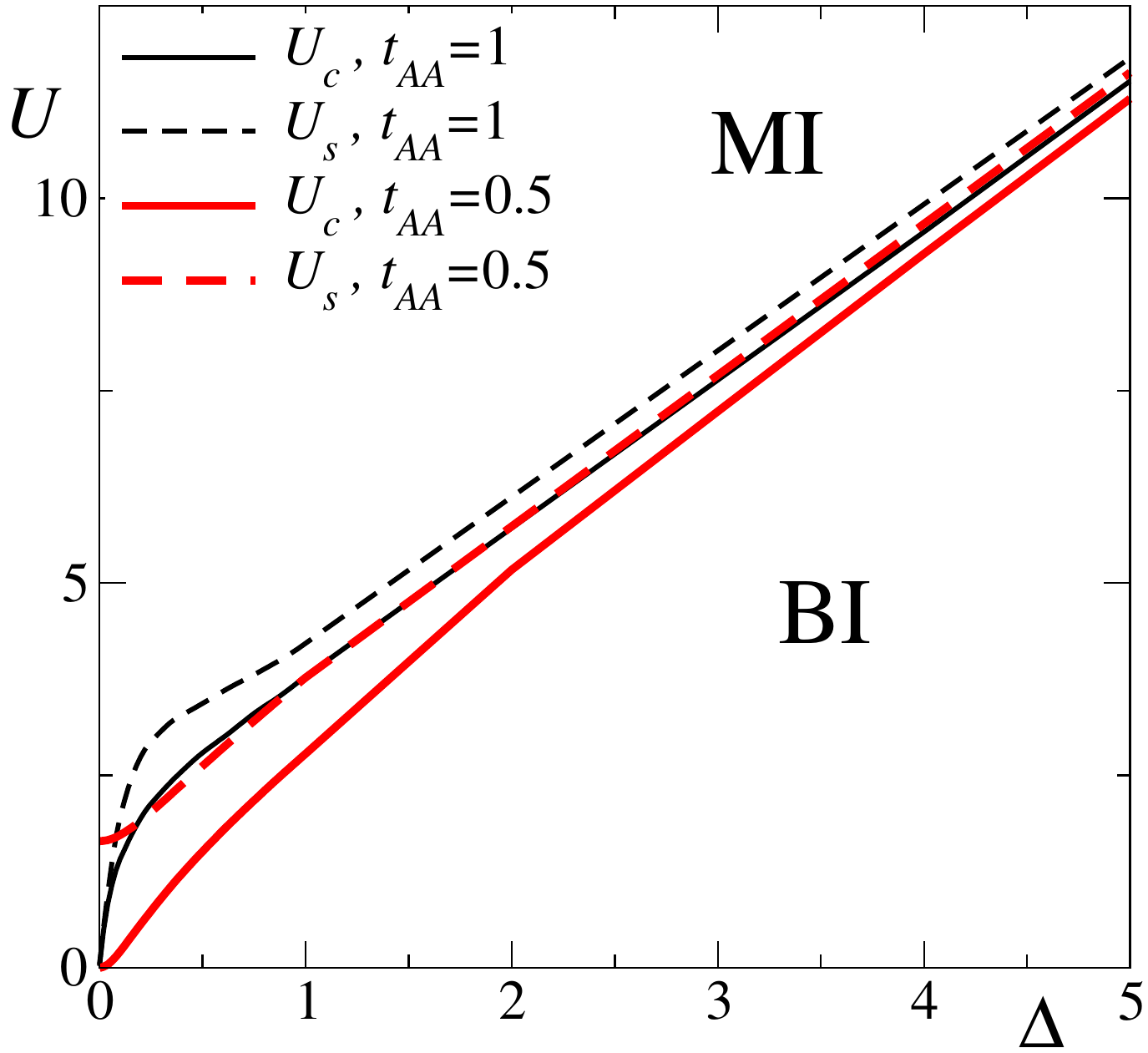}
\caption{(Color online) Phase diagram of the IHM with DDH in the 
$\Delta, U$ plane
for $t_{AB}=1$, and two values of $t_{AA}=t_{BB}$. The region between the full and dashes lines corresponds to the SDI.}
\label{larged}
\end{center}
\end{figure}

In Fig. \ref{larged}, we compare the phase diagram of the
standard IHM with that in which the hopping terms that do not 
alter the total number of singly occupied sites $t_{AA}=t_{BB}$ is
reduced. For fixed $\Delta$ the system is a BI for low $U$ 
and a MI for large $U$. Both phases are separated by a narrow 
region of the SDI phase. Increasing $U$, the charge 
transition at $U=U_c$ 
(with a jump in $\gamma_c$ from 0 to $\pi$ \cite{phihm}) corresponds to the change from the BI to 
the SDI,
and at the spin transition for $U=U_s$ (with a jump in $\gamma_c$ from 0 to $\pi$ \cite{phihm}) the SDI changes to the MI.  

For $\Delta > \sim 3 t_{AB}$ the width of the SDI phase 
is of the order of a fraction of $t_{AB}$. Naturally,
keeping the three hopping terms equal $t_{\alpha \beta}=t$ and reducing $t$ the SDI phase shrinks and both 
$U_c,U_s \rightarrow 2 \Delta$ for $t \rightarrow 0$. 
It is therefore noticeable that reducing only $t_{AA}=t_{BB}$,
the extension of the SDI phase is {\em increased} 
and by about 20\% for $\Delta > 3 t_{AB}$.

\begin{figure}[h]
\begin{center}
\includegraphics[width=\columnwidth]{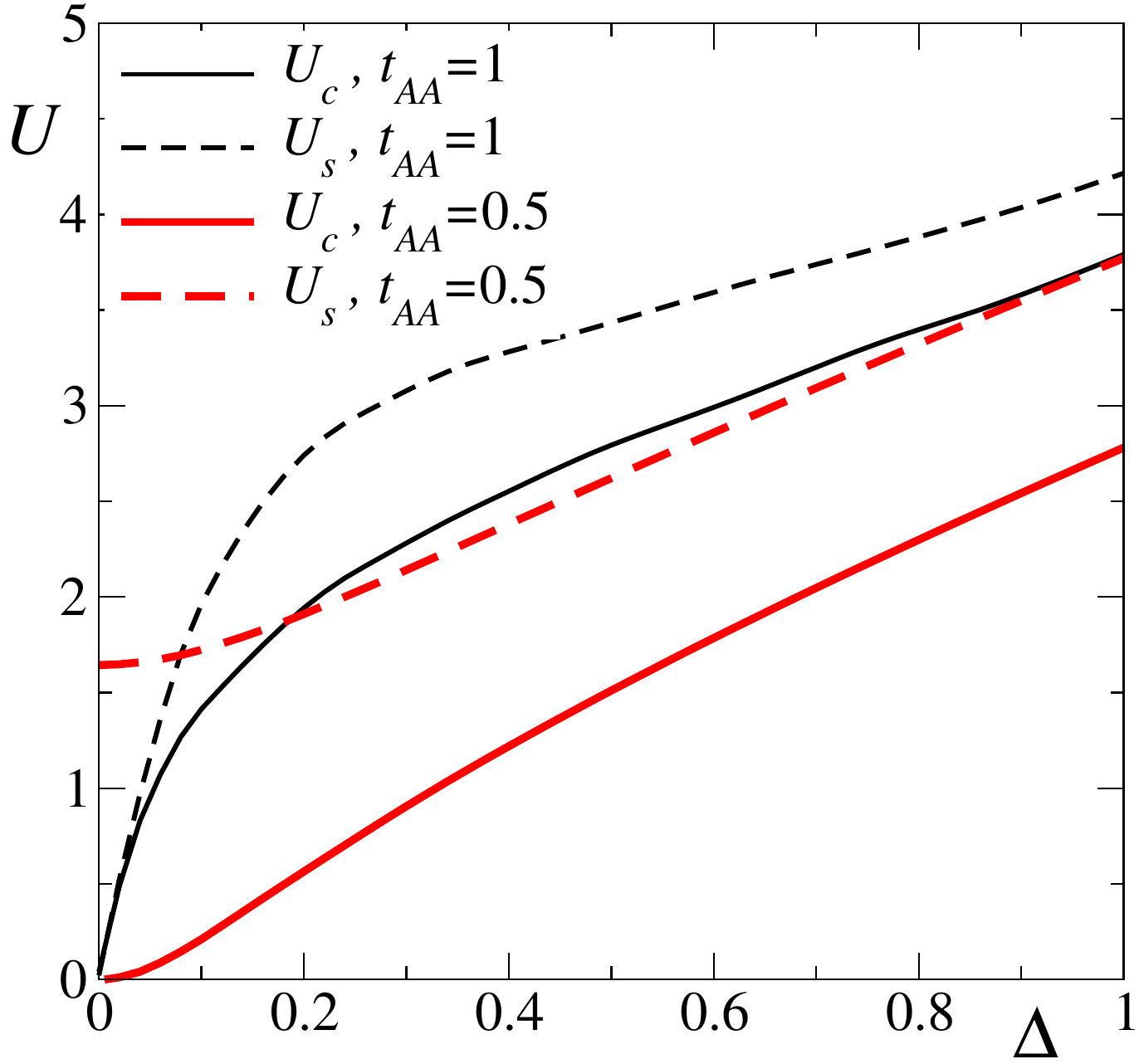}
\caption{Same as Fig. \ref{larged} in a smaller region of 
$\Delta$.}
\label{smalld}
\end{center}
\end{figure}

As it is apparent in Fig. \ref{smalld}, this effect is more dramatic for $\Delta < 0.5 t_{AB}$. In fact, contrary to the case 
of equal $t_{\alpha \beta}=t$, there is a finite spin gap for small 
$U$ even at $\Delta=0$ when $t_{AB} > t_{AA}=t_{BB}$. This result
has been found before \cite{jaka,bosolili,phtopo,naka2} and can be
understood from analytical calculations using bosonization 
\cite{jaka,bosolili} which coincide very well with numerical
calculations \cite{phtopo} for small vales of $U$.

The particular features of the phase diagram for small $\Delta$, 
render it possible to perform time evolutions around a critical point for the charge transition that transport a quantized unit of charge per cycle with open charge and spin gaps in the whole cycle.
For example for  $\Delta=0.3$, $t_{AB}=1$, and $t_{AA}=t_{BB}=0.5$, 
we find $U_c=0.91$ and $U_s=2.14$. Similarly 
for  $\Delta=0.4$ we find $U_c=1.22$ and $U_s=2.38$.
Performing a time dependent cycle in the plane either $(\delta, \Delta)$
or $(\delta, U)$ with center at the charge critical point 
(with $\delta=0$) and amplitude in $\Delta$ of about $\pm 0.25$ 
or in $U$ near $\pm 0.5$, the cycle never reaches the MI phase 
and therefore, the spin gap is always open. 
One point that should be taken into account is that the spin transition is of the Kosterlitz-Thouless type, and therefore the spin gap is exponentially small in the SDI phase near the transition boundary \cite{phtopo}. Therefore it might be convenient to move the time cycle away from the MI-SDI boundary, keeping the critical point inside it.

\begin{figure}[h]
\begin{center}
\includegraphics[width=\columnwidth]{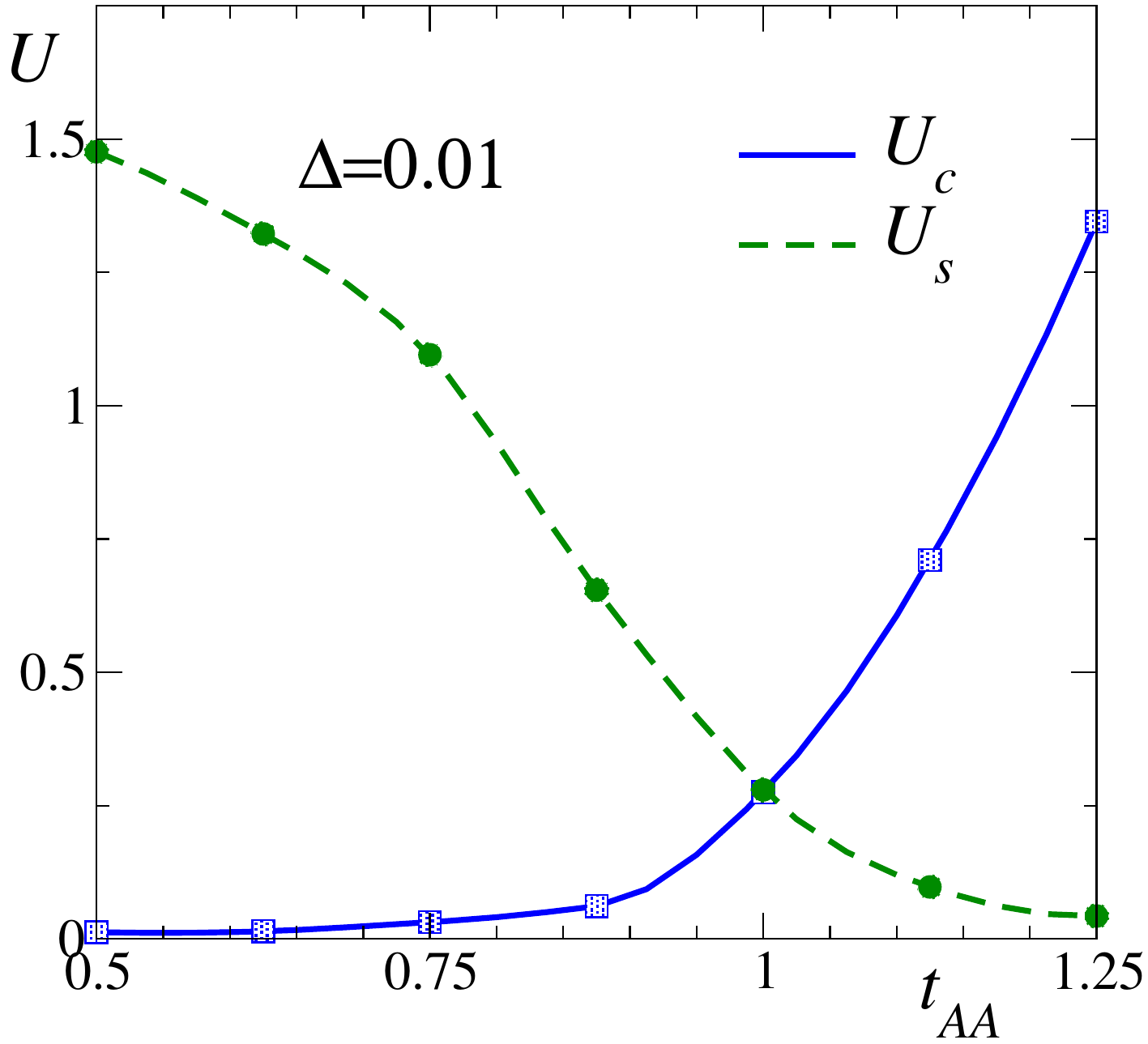}
\caption{(Color online) Critical values of $U$ at the charge and spin transitions as a function of $t_{AA}=t_{BB}=$ for 
$t_{AB}=1$ and $\Delta=0.01$.}
\label{uvstaa}
\end{center}
\end{figure}

In the previous figures we have taken $t_{AA}=t_{BB}=t_{AB}/2$.
In Fig. \ref{uvstaa} we show how the values of $U$ at both transitions change with $t_{AA}=t_{BB}$ for a small value 
of $\Delta$. We can see that the 
change is more rapid for $t_{AA}$ near $t_{AB}$ and the
increase in $U_s$ is already large for $t_{AA}/t_{AB}=3/4$.
Note also that when $t_{AA}/t_{AB}$ exceeds 1 for a significant
amount $U_c$ becomes larger than $U_s$ giving rise to a new 
phase in between. The properties of this phase are beyond the 
scope of the present work. For $\Delta=0$, this phase corresponds
to a Tomonaga-Luttinger liquid with triplet superconducting and bond spin-density wave correlations dominating at large distances
\cite{jaka,bosolili,phtopo}, but $\Delta$ is a relevant perturbation that modifies the physics.

\section{Summary and discussion}

\label{sum} 

We have calculated the quantum phase diagram of the IHM including 
electron-hole symmetric DDH, which corresponds to Eq. (\ref{hirm})
with $\delta=0$ and $t_{AA}=t_{BB} < t_{AB}$, using the MCEL
in rings of up to 14 sites. 
We obtain that a reduction of $t_{AA}=t_{BB}$ with respect to $t_{AB}$, increases the region of the phase diagram occupied
by the fully gapped SDI phase, particularly for $|\Delta| <t_{AB}$
and $U < 2 t_{AB}$. 

This result is of possible relevance
to experiments with cold atoms for which quantized pumping
is observed, but crossing the spin gapless MI phase leads
to oscillation and the breakdown of topological pumping after
the first cycle. Floquet engineering renders it possible to 
achieve the region $t_{AA}=t_{BB} < t_{AB}$ and enlarge the 
region of the gapped SDI phase. 
To confirm the possibilities of this proposal, it would be useful
to calculate the spin gap and the internal gap between even and odd singlets in the SDI phase. This would require a study of longer chains using density-matrix renormalization group.
It would also be useful to simulate the time dependence 
in pumping cycles similar to the ones suggested here, using
infinite time-evolving block decimation.

\section*{Acknowledgments}

We thank Konrad Viebahn, Eric Bertok and Fabian Heidrich-Meisner for useful discussions.  AAA acknowledges financial support provided by PICT 2017-2726 and PICT 2018-01546
of the ANPCyT, Argentina.


\end{document}